\documentclass[%
 reprint,
groupedaddress,
preprintnumbers,
nofootinbib,
amsmath,amssymb,
aps,
eqsecnum,
notitlepage
]{revtex4-2}

\usepackage{graphicx}
\usepackage{dcolumn}
\usepackage{bm}
\usepackage{hyperref}
\usepackage[subrefformat=parens]{subcaption}
\usepackage{booktabs} 
\usepackage{orcidlink}

\usepackage{url}
\hypersetup{colorlinks=true, citecolor=cyan}

\newcommand{\be}{\begin{equation}}
\newcommand{\ee}{\end{equation}}

\usepackage{color}
\definecolor{mygreen}{RGB}{0,130,0} 

\newcommand{\aye}{\mathrm{i}}

\usepackage{comment}

\begin{document}

\preprint{RESCEU-2/25}

\title{Statistical biases in parametrized searches for gravitational-wave polarizations}

\author{Hayato Imafuku${}^{1,2}$
\orcidlink{0009-0001-3490-8063}}
\author{Hiroki Takeda${}^{3,4}$
\orcidlink{0000-0001-9937-2557}}
\author{Atsushi Nishizawa${}^{5,6,2}$
\orcidlink{0000-0003-3562-0990}}
\author{Daiki Watarai${}^{1,2}$
\orcidlink{0009-0002-7569-5823}}
\author{Kipp Cannon${}^{2}$
\orcidlink{0000-0003-4068-6572}}

\affiliation{${}^1$Graduate School of Science, The University of Tokyo, Tokyo 113-0033, Japan}
\affiliation{${}^2$Research Center for the Early Universe (RESCEU), Graduate School of Science, The University of Tokyo, Tokyo 113-0033, Japan}
\affiliation{${}^3$The Hakubi Center for Advanced Research, Kyoto University, Yoshida Ushinomiyacho, Sakyo-ku, Kyoto 606-8501, Japan}
\affiliation{${}^4$Department of Physics, Kyoto University, Kyoto 606-8502, Japan}
\affiliation{${}^5$Physis Program, Graduate School of Advanced Science and Engineering, Hiroshima University, Higashi-Hiroshima, Hiroshima 739-8526, Japan}
\affiliation{${}^6$Astrophysical Science Center, Hiroshima University, Higashi-Hiroshima, Hiroshima 739-8526, Japan}

\date{\today}

\begin{abstract}
In tests of gravity using gravitational waves (GWs), GW events analyzed are often selected based on specific criteria, particularly the signal-to-noise ratio. However, such event selection can introduce bias into parameter estimation unless the selection effect is appropriately taken into account in the analysis. In this paper, we investigate how event selection with certain prior information affects parameter inference within the scalar-tensor polarization framework, focusing on the measurement of the scalar mode amplitude parameters. We find that for the Tensor+Scalar(dipole) model, the amplitude of the scalar dipole radiation is overestimated when its true value is nonzero while there is no false deviation in the absence of the scalar mode. The same bias is expected to occur also for the Tensor+Scalar(quadrupole) model. However, the error typically exceeds the bias as the scalar quadrupole mode is difficult to be distinguished from the tensor mode.
\end{abstract}

\maketitle


\section{introduction}
\label{sec:introduction}
The standard theory of gravity, general relativity (GR), is consistent with past experiments and observations (e.g.,~\cite{Will:2014kxa, Will_2018}). However, the majority of these tests have been carried out in weak gravitational fields, such as those within the solar system. The first observation of gravitational waves (GWs) in 2015 opened a new window for testing GR in strong gravitational fields~\cite{LIGOScientific:2016aoc, LIGOScientific:2016lio}. Since then, multiple GW signals from compact binary coalescences (CBCs) have been detected by LIGO-Virgo-KAGRA (LVK) collaboration~\cite{LIGOScientific:2018mvr, LIGOScientific:2020ibl, KAGRA:2021vkt}. To date, GR has undergone extensive testing in strong gravitational fields (e.g.,~\cite{LIGOScientific:2019fpa, LIGOScientific:2020tif, LIGOScientific:2021sio}), and no clear violation has been identified with the current level of observational precision.

The search for GW polarizations is a powerful tool for testing gravity. In GR, GWs consist of only two tensor modes: the plus mode and the cross mode. However, some modified theories of gravity predict that GWs can have the scalar polarization mode in addition to the tensor modes due to the existence of additional scalar degrees of freedom~\cite{Eardley:1973br, PhysRevD.8.3308}. For instance, in Einstein-dilaton Gauss-Bonnet gravity and dynamical Chern-Simons gravity, GWs emitted from binary black holes (BBHs) may contain scalar modes, as BHs can possess scalar charges~\cite{Berti:2018cxi, Tahura:2018zuq}. The number and types of GW polarizations reflect the degrees of freedom inherent to a gravitational field. Therefore, exploring the polarizations allows us to test GR with a focus on the degrees of freedom of the gravitational field and to approach the fundamental nature of gravity. The discovery of evidence for nontensor modes would signify a violation of GR, necessitating modifications to the theory.

The polarization modes of GWs have been investigated through various methods. One category of tests that does not require a waveform model includes null-stream tests~\cite{Chatziioannou:2012rf,LIGOScientific:2021sio, Pang:2020pfz, Hagihara:2019ihn} and wavelet analysis \cite{Chatziioannou:2021mij}. These results predominantly support the GR hypothesis. This method allows for hypothesis comparison but has less sensitivity. On the other hand, some studies assume theoretical waveforms. In pure polarization tests~\cite{LIGOScientific:2017ycc, LIGOScientific:2018dkp, Takeda:2020tjj}, the hypotheses that GWs possess only the tensor, the vector, or the scalar mode are evaluated. The results strongly prefer the case of pure tensor modes over the cases of pure scalar or vector modes. However, most modified gravity theories predict a mixture of the dominant tensor mode and subdominant nontensor modes. The mixed polarization tests with a parametrized scalar-tensor waveform model were performed in~\cite{Takeda:2021hgo, Takeda:2023wqn}. By parametrizing non-GR effects, we can directly constrain non-GR parameters with high sensitivity and interpret the physical implications of the results more clearly. The results are consistent with GR; however, their precision is limited to approximately $\sim\mathcal{O}(10^{-1})$, compared to the tensor amplitude.

A straightforward extrapolation of the observational constraints on gravity theories in the weak field regime to the strong regime shows that the scalar mode amplitude in the observed GW signal is suppressed to at least $\mathcal{O}(10^{-2.5})$ relative to the tensor mode amplitude~\cite{Takeda:2023mhl}. Therefore, a statistical search involving multiple GW signals is required to achieve higher sensitivity in testing. However, when testing GR using GWs, analyzed signals are often selected based on specific criteria, such as the signal-to-noise ratio (SNR), the detector network, and the source properties. Such event selections, in general, introduce parameter estimation biases, which we call statistical biases, unless the effect is properly accounted for in the analysis process.

The process of selecting preferred events for analysis creates a mismatch between the statistical properties of the ensemble of events used for the analysis and those of the astrophysical population from which they have been drawn, especially for the luminosity distance.
For example, selecting for high SNR introduces a strong correlation between the luminosity distance and orbital inclination of the source (e.g.,~\cite{Cutler:1994ys, Nissanke:2009kt, Usman:2018imj, Veitch:2012df, Narikawa:2017wsz}), which is a property that is not possessed by the natural population of GW sources.  Inferring the properties of observed sources without accounting for this induced correlation leads to a bias in the estimation of the inclination angle.

In this paper, we investigate a possible bias in the estimation of the scalar mode amplitude when particular events are selected for the analysis.
The measurement of the scalar mode amplitude can be affected by the event selection through the bias on the inferred inclination angle mentioned above, because each polarization exhibits a distinct angular pattern of GW radiation, characterized by a unique inclination-angle dependence~\cite{Takeda:2018uai}.
Section~\ref{sec:parametrized scalar tensor waveform model} presents the construction of a parametrized scalar-tensor waveform model by introducing non-GR parameters.
In Sec.~\ref{sec:Parameter estimation method}, we perform injection searches. First, we investigate the recovery of the inclination angle in pure polarization cases to see the tendency of the bias for each mode individually. Second, based on those results, we evaluate the effect of event selection on parameter estimation for both the inclination angle and the scalar mode amplitude in mixed polarization cases.
The results of each injection search are presented in Sec.~\ref{sec:Results}.
Finally, Sec.~\ref{sec:Discussions and Conclusions} is devoted to the discussions and conclusions, with an outlook for future work. Throughout this paper, we adopt geometric units, setting $G=c=1$.

\section{parametrized scalar-tensor waveform model}
\label{sec:parametrized scalar tensor waveform model}
In this section, we outline a parametrized framework for scalar-tensor inspiral GWs from compact binary coalescences, following~\cite{Takeda:2023wqn, Takeda:2021hgo}. In general, there are two scalar modes: the breathing mode and the longitudinal mode. However, since the antenna pattern functions of the interferometric detector are degenerate for these two modes, the framework adopts only the breathing mode as the third polarization. Thus, the waveform model consists of two tensor modes and the scalar breathing mode. 

We begin by assuming the $\ell$th term in the multipole expansion, expressed as follows~\cite{Chatziioannou:2012rf},
\begin{align}
    h^{(\ell)}_{p}(t) = \frac{1}{2}g^{(\ell)}_{p}(\iota)\eta^{(2-\ell)/5}\frac{4\mathcal{M}}{d_L}\left(2\pi\mathcal{MF}\right)^{\ell/3}e^{+\aye\ell\Phi}\:,
    \label{eq:l-th TD WF}
\end{align}
where $p$ runs over indices of each polarization, plus ($+$), cross ($\times$), and breathing ($b$) mode, and $g^{(\ell)}_{p}(\iota)$ is a function of inclination angle $\iota$, which describes the angular pattern of GW radiation~\cite{Takeda:2018uai}. In Eq.~\eqref{eq:l-th TD WF}, $\mathcal{M}$ is the redshifted chirp mass, $d_L$ is the luminosity distance, $\eta$ is the symmetric mass ratio, $\mathcal{F}$ is the orbital frequency, and $\Phi$ is the orbital phase. We consider the multipoles up to quadrupole. Hence, the harmonic index is $\ell=2$ for the tensor modes and $\ell=1,2$ for the scalar dipole and quadrupole modes, respectively. For these modes, we can write down the polarization components explicitly,
\begin{align}
    h^{(2)}_{+}(t)&=-\frac{1}{2}\frac{1+\cos^2\iota }{2}\frac{4\mathcal{M}}{d_L}\left(2\pi\mathcal{MF}\right)^{2/3} e^{+\aye2\Phi} \:, \label{eq:TD plus} \\
    h^{(2)}_{\times}(t)&=-\frac{1}{2}i\cos\iota\frac{4\mathcal{M}}{d_L}\left(2\pi\mathcal{MF}\right)^{2/3} e^{+\aye2\Phi}\:, \label{eq:TD cross} \\
    h^{(1)}_{b}(t)&=\frac{1}{2}A_{b1}\sin \iota\,\eta^{1/5}\frac{4\mathcal{M}}{d_L}\left(2\pi\mathcal{MF}\right)^{1/3} e^{+\aye\Phi}\:, \label{eq:TD breathing dipole} \\
    h^{(2)}_{b}(t)&=\frac{1}{2}A_{b2}\sin^2\iota \frac{4\mathcal{M}}{d_L}\left(2\pi\mathcal{MF}\right)^{2/3} e^{+\aye2\Phi}\:. \label{eq:TD breathing quad}
\end{align}
Here, we introduce new parameters, $A_{b1}$ and $A_{b2}$, which characterize the amplitude of the scalar dipole and quadrupole radiation, respectively.

Furthermore, we modify the stress-energy tensor by introducing a coupling parameter $\gamma$ to account for the coupling strength between a scalar field and metric. Using Eqs.~\eqref{eq:TD plus}--\eqref{eq:TD breathing quad}, we calculate the GW energy flux, which is corrected by the additional scalar radiation as
\begin{align}
    \dot{E}_{\mathrm{GW}} &= - \frac{d^{2}_{L}}{16\pi} \int{\mathrm d}\Omega \left\{ \langle \dot{h}^{2}_{+} + \dot{h}^{2}_{\times} \rangle + \gamma \langle \dot h^{2}_{b1} + \dot{h}^{2}_{b2} \rangle \right\} \nonumber \\
    &= \dot{E}_{\rm{GR}} \left(1 + \frac{5}{24} \gamma A^{2}_{b1} \left(2\pi\mathcal{MF}\right)^{-2} \eta^{2/5} + \frac{2}{3} \gamma A^{2}_{b2} \right)\:, \label{eq:modified enery flux}
\end{align}
where $\dot{E}_{\mathrm{GR}}$ represents the GW energy flux in GR. The symbol $\langle\cdots\rangle$ denotes averaging over several periods of GWs. From Eq.~(\ref{eq:modified enery flux}), the Fourier transform of the GW signal in the stationary phase approximation~\cite{Maggiore:2007ulw, Droz:1999qx, Yunes:2009yz} is expressed as the sum of each polarization
\begin{align}
    \tilde{h}_{I}(f) = \tilde{h}^{(2)}_{+}(f) + \tilde{h}^{(2)}_{\times}(f) + \tilde{h}^{(1)}_{b}(f) + \tilde{h}^{(2)}_{b}(f)\:, \label{FD:full signal}
\end{align}
where
\begin{align}
    \tilde{h}^{(2)}_{+}(f) &= -F^{+}_{I}\sqrt{\frac{5\pi}{96}}\,\frac{\mathcal{M}^{2}}{d_L} (1+\cos^2 \iota)\,u_{2}^{-7/2} \nonumber \\
    &\qquad \times \left[ 1+\delta A^{(2)} \right] e^{-\aye \Psi^{(2)}_{\mathrm{GR}}}e^{-\aye \delta\Psi^{(2)}}\:, \label{eq:FD plus} \\
    \tilde{h}^{(2)}_{\times}(f) &= F^{\times}_{I}\sqrt{\frac{5\pi}{96}}\,\frac{\mathcal{M}^{2}}{d_L} (2i\cos\iota) \,u_{2}^{-7/2} \nonumber \\
    &\qquad \times \left[ 1+\delta A^{(2)} \right]  e^{-\aye \Psi^{(2)}_{\mathrm{GR}}}e^{-\aye \delta\Psi^{(2)}}\:, \label{eq:FD cross} \\
    \tilde{h}^{(1)}_{b}(f) &=F^{b}_{I}\sqrt{\frac{5\pi}{48}}A_{b1} \frac{\mathcal{M}^2}{d_L} \eta^{1/5}(2 \sin \iota)u_{1}^{-9/2} \nonumber \\
    &\qquad \times e^{-\aye\Psi^{(1)}_{\mathrm{GR}}}e^{-\aye \delta\Psi^{(1)}}\:, \label{eq:FD dipole} \\
    \tilde{h}^{(2)}_{b}(f) &= F^{b}_{I}\sqrt{\frac{5\pi}{96}} A_{b2} \frac{\mathcal{M}^2}{d_L} (2\sin^2 \iota)u_{2}^{-7/2} \nonumber \\
    &\qquad \times e^{-\aye \Psi^{(2)}_{\mathrm{GR}}}e^{-\aye \delta\Psi^{(2)}}\:. \label{eq:FD quadru}
\end{align}
Here, $\tilde{h}^{(2)}_{+}(f)$, $\tilde{h}^{(2)}_{\times}(f)$, $\tilde{h}^{(1)}_{b}(f)$, and $\tilde{h}^{(2)}_{b}(f)$ correspond to the quadrupole radiation of the plus mode and the cross mode, and the dipole and quadrupole radiation of the scalar breathing mode, respectively. The functions $F^{A}_{I}$ with $A=\{+,\times,b\}$ are the antenna pattern functions for the $I$th detector, which depend on sky localization and the polarization angle~\cite{Nishizawa:2009bf}. In addition, $\Psi^{(\ell)}_{\mathrm{GR}}$ represents the GW phase of the $\ell$th harmonic mode defined by
\begin{align}
    \Psi^{(\ell)}_{\mathrm{GR}} = 2\pi ft_c - \ell\Phi_c - \frac{\pi}{4} + \frac{3\ell}{256}u^{-5}_{\ell}\sum^{7}_{i=0}\phi_{i}u^{i}_{\ell} \:,
\end{align}
although only the $\ell=2$ mode exists in GR. Here, $t_c$ and $\Phi_c$ are the coalescence time and phase, respectively. $\phi_i$ are post-Newtonian coefficients~\cite{Khan:2015jqa}, and the reduced $\ell$th harmonic frequency $u_{\ell}$ is defined as
\begin{align}
    u_\ell = \left(\frac{2\pi\mathcal{M}f}{\ell}\right)^{1/3}\:,
\end{align}
where $f=\ell\mathcal{F}$ is the GW frequency. The correction terms $\delta A^{(\ell)}$ and $\delta\Psi^{(\ell)}$ account for the amplitude and phase modifications due to the scalar mode backreaction, and are given by~\cite{Takeda:2023wqn}
\begin{align}
    \delta A^{(\ell)} &= -\frac{5}{48}\tilde{A}^{2}_{b1}\eta^{2/5}u_{\ell}^{-2} -\frac{1}{3}\tilde{A}^{2}_{b2}\:, \\
    \delta \Psi^{(\ell)} &= -\frac{5\ell}{3584} \tilde{A}^{2}_{b1} \eta^{2/5} u_{\ell}^{-7} - \frac{\ell}{128} \tilde{A}^{2}_{b2} u_{\ell}^{-5}\:,
\end{align}
where
\begin{align}
    \tilde{A}_{b1} &= \sqrt{\gamma}A_{b1}\:, \\
    \tilde{A}_{b2} &= \sqrt{\gamma}A_{b2}\:.
\end{align}
In these calculations, we include terms up to the second order in $A_{b1}$ and $A_{b2}$.

Finally, we obtain the modified inspiral GW waveform, which contains the two tensor modes and one scalar mode, as shown in Eqs.~\eqref{eq:FD plus}--\eqref{eq:FD quadru}. We assume generic modifications to the stress-energy tensor and the GW energy flux, Eq.~\eqref{eq:modified enery flux}, without relying on a specific theory. This model introduces four additional parameters: $A_{b1}$ and $A_{b2}$, which characterize the scalar mode amplitude, and $\tilde{A}_{b1}$ and $\tilde{A}_{b2}$, which contribute to both the amplitude and phase correction terms. Note that these four parameters correspond to the non-GR parameters appearing in the parametrized post-Einsteinian (ppE) framework~\cite{Yunes:2009ke, Cornish:2011ys, Chatziioannou:2012rf, Tahura:2018zuq}. This parametrization, rather than one with $\gamma$, enables a more straightforward comparison of the correction terms with the ppE parameters. The parametrized scalar-tensor waveform enables a systematic search for the scalar polarization of GWs, covering many modified theories of gravity.

\section{Parameter estimation method}
\label{sec:Parameter estimation method}
The primary objective here is to investigate the parameter estimation biases in the polarization tests caused by selecting only loud GW signals, using the parametrized waveform constructed in Sec.~\ref{sec:parametrized scalar tensor waveform model}. The scalar-tensor waveform incorporates four independent non-GR parameters. Among them, $A_{b1}$ and $A_{b2}$ appear in the waveform amplitude, as well as the other amplitude parameters such as the luminosity distance and the inclination angle. It is anticipated that $A_{b1}$ and $A_{b2}$ would correlate with the other amplitude parameters and the estimation of these parameters would be affected by the mismatch between the intrinsic source distribution and the prior distribution for luminosity distance.

Our analysis is based on Bayesian inference, which computes the posterior probability distribution $p(\bm{\theta}|\bm{d},M)$ according to the Bayes' theorem~\cite{BayesLIIAE, Maggiore:2007ulw, LIGOScientific:2019hgc},
\begin{align}
    p(\bm{\theta}|\bm{d},M) = \frac{p(\bm{\theta}|M)p(\bm{d}|\bm{\theta},M)}{p(\bm{d}|M)}\:,
\end{align}
where $p(\bm{\theta}|M)$ is the prior probability distribution, $p(\bm{\theta}|\bm{d},M)$ is the likelihood, and $p(\bm{d}|M)$ is the evidence. $M$ represents the hypothetical model, which in this case corresponds to the scalar-tensor waveform model. $\bm{d}=\{d_I\}^{N}_{I=1}$ is the set of the GW strain data observed by each detector labeled by $I$, and $\bm{\theta}$ is the set of the waveform model parameters given by the standard 11 source  parameters for an aligned-spin binary along with the additional two non-GR parameters. Those are given by
\begin{align}
    \bm{\theta} &= \{m_{1,\mathrm{source}},m_{2,\mathrm{source}},\chi_1,\chi_2,d_L,\iota,\alpha,\delta,\psi,\phi_{\mathrm{ref}},t_c, \nonumber \\
    &\qquad A_{b1},\tilde{A}_{b1}\}
\end{align}
for the Tensor+Scalar(dipole) model, and
\begin{align}
    \bm{\theta} &= \{m_{1,\mathrm{source}},m_{2,\mathrm{source}},\chi_1,\chi_2,d_L,\iota,\alpha,\delta,\psi,\phi_{\mathrm{ref}},t_c, \nonumber \\
    &\qquad A_{b2},\tilde{A}_{b2}\}
\end{align}
for the Tensor+Scalar(quadrupole). Here $m_{1,\mathrm{source}}$ and $m_{2,\mathrm{source}}$ are the source-frame masses of the heavier and lighter objects, respectively. Each object has a dimensionless spin $\chi_1$ and $\chi_2$ assuming  an aligned-spin binary. $\alpha$ and $\delta$ are the right ascension and declination, respectively. Additionally, $\psi$ is the polarization angle, $\phi_{\mathrm{ref}}$ is the phase at the reference frequency $f_{\mathrm{ref}}=20~\mathrm{Hz}$, and $t_c$ is the coalescence time. Note that in the following, the mass parameter with the notation of ``source'' refers to a source-frame value, and without it corresponds to a detector-frame value.

The likelihood is assumed to be a Gaussian noise likelihood, computed under the assumption that the detector noise is stationary and Gaussian. It is expressed as
\begin{align}
    p(\bm{d}|\bm{\theta},M) \propto \exp\left\{-\frac{1}{2}\sum_{I}\langle d_{I}-h_{I}(\bm{\theta})|d_{I}-h_{I}(\bm{\theta})\rangle\right\}\:,
\end{align}
where the angle bracket $\langle\cdots|\cdots \rangle$ denotes the noise-weighted inner product defined by
\begin{align}
    \langle a|b \rangle = 4\mathrm{Re}\int^{f_{\mathrm{max}}}_{f_{\mathrm{min}}} \frac{a^{*}(f)b(f)}{S_{n,I}(f)} \mathrm{d}f\:.
\end{align}
Here, $S_{n,I}(f)$ is the noise power spectral density and $I$ labels the detector. $I$ runs over three detectors, $I=$ \{LIGO Hanford, LIGO Livingston, Virgo\}. The power spectral densities adapted are the O4 design sensitivity of each detector~\cite{Capote:2024rmo, Virgo:2019juy}. The lower cutoff frequency $f_{\mathrm{min}}$ is set to 20 Hz, while the upper cutoff frequency $f_{\mathrm{max}}$ is set to the innermost stable circular orbit (ISCO) frequency for the Schwarzschild BH. The upper cutoff frequency is given by $f_{\mathrm{max}}=(6\sqrt{6}\pi(m_1+m_2))^{-1}$, since our theoretical waveform is valid only for the inspiral phase. Note that, in general, the upper cutoff frequency can vary depending on the spins of the binary components and the underlying theories of gravity. However, the Schwarzschild ISCO frequency is commonly adapted as the upper cutoff frequency under the assumption that it is predominantly determined by the total mass.  For simplicity, we adopt this approach in our analysis.

We perform Bayesian inference using the Bilby software~\cite{Ashton:2018jfp} and the dynamic Nested Sampling~\cite{Higson_2018, Speagle:2019ivv}. The sampler settings are based on Ref.~\cite{Romero-Shaw:2020owr}. Specifically, we set the number of live points (nlive) to 1500, the sampling method to acceptance-walk and the number of walks to 200. Note that we confirmed that this nlive value is sufficient to ensure convergence of the sampling. For the GR template waveform of the inspiral phase, we use TaylorF2~\cite{Buonanno:2009zt}, as implemented in the LVK Algorithm Library---LALSuite~\cite{lalsuite}. The noise for each detector is a random Gaussian realization created using its respective noise power spectral density. In the analysis, we assume the flat $\Lambda$ cold dark matter cosmological model with parameters given by Planck15~\cite{Planck:2015fie}.

We consider two cases: the pure polarization case and the mixed polarization case. To assess the effect of event selection on parameter estimation for the amplitude parameters of each mode, we first investigate the recovery of the inclination angle for the pure tensor case and the pure scalar case individually as a guide for the mixed polarization cases considered later. 

Here, the pure tensor (scalar) polarization case refers to GW signals containing only the tensor (scalar) polarization. In this analysis, we inject 50 GW signals for each inclination angle $\iota=\pi/6, \pi/4, \pi/3, \pi/2$ using the parameter sets shown in Table~\ref{tab:injected values for inclination recovery}. The right ascension $\alpha$ and the declination $\delta$ are sampled uniformly on the sky to mitigate the dependence on the antenna pattern function, as our goal is to evaluate the effects of event selection and the correlation between the inclination angle and the luminosity distance. We fix the spin parameters to zero for both the injection and template waveforms, because the parameters that we want to estimate are those at the Newtonian order in amplitude and are not significantly affected by spins in phase. For simplicity, the amplitude and phase correction terms associated with $\tilde{A}_{b1}$ and $\tilde{A}_{b2}$ are excluded, as we are solely focused on recovering the inclination angle in each pure polarization case, thus, $\tilde{A}_{b1}=\tilde{A}_{b2}=0$. This simplification does not affect the following results, since $\tilde{A}_{b1}$ and $\tilde{A}_{b2}$ are primarily determined by the phase term, whereas the inclination angle is estimated by the amplitude term.

The priors on the GR parameters are applied as listed in Table~\ref{tab:priors}. Note that we fix $\chi_{1,2}=0$ by applying the delta function prior for the simplicity of calculations, as mentioned in the previous paragraph.  For the additional non-GR parameters, we adopt uniform priors within the range $[-1,1]$ under the assumption that the magnitude of the scalar mode amplitude is smaller than that of the tensor mode amplitude. Indeed, Eqs.~\eqref{eq:FD plus}--\eqref{eq:FD quadru} are derived by adapting this assumption.
The prior for the luminosity distance is employed by the uniform distribution in comoving volume and source-frame time.

For each injection, the parameter sets in Table~\ref{tab:injected values for inclination recovery} give SNRs in the range of approximately $20-100$. Consequently, the distributions of these 50 GW signals do not follow the prior distribution shown in Table~\ref{tab:priors}.
\begin{table}[ht]
  \caption{The injected values for the pure polarization case.}
  \centering
    \begin{tabular}{*{2}{wc{40mm}}}
      \toprule
      \multicolumn{1}{c}{Parameter} & \multicolumn{1}{c}{Injected values} \\
      \midrule
      $m_{1,\mathrm{source}}$ & $15~M_{\odot}$ \\
      $m_{2,\mathrm{source}}$ & $10~M_{\odot}$ \\
      $\chi_{1,2}$ & $0$ \\
      $d_{L}$ & $204~\mathrm{Mpc}$ \\
      $\iota$ & $\{\pi/6, \pi/4, \pi/3, \pi/2\}$ \\
      $\alpha$ & sampled uniformly \\
      $\sin\delta$ & sampled uniformly \\
      $\psi$ & $2.659~\mathrm{rad}$ \\
      $\phi_{\mathrm{ref}}$ & $1.3~\mathrm{rad}$ \\
      $t_{c}$ & $186741861.5~\mathrm{s}$ \\
      \bottomrule
    \end{tabular}
  \label{tab:injected values for inclination recovery}
\end{table}

In the mixed polarization case, we focus on measuring the inclination angle and the scalar mode amplitude by considering both the Tensor+Scalar(dipole) model and the Tensor+Scalar(quadrupole) model, as the injection and template waveform. We inject GW signals three times for each scalar mode amplitude, setting $A_{b1} (A_{b2})=0,0.1,0.3,0.5$, while fixing $A_{b2} (A_{b1})=0$ and removing $A_{b2} (A_{b1})$ from the estimated parameters. The remaining parameters are identical to those used for the pure polarization case shown in Table~\ref{tab:injected values for inclination recovery}, except for the right ascension and declination, whose injection values are set to $\alpha=0.833~\mathrm{rad}$ and $\delta=-0.784~\mathrm{rad}$. For ease of computation, we adopt the single ra-dec set in this case. This choice does not yield any distinctive or biased results, as shown in Appendix~\ref{sec:Validity of ra-dec choice for mixed polarization}. In addition, we choose $\tilde{A}_{b1}=\tilde{A}_{b2}=0$, and these parameters are estimated in the analysis in contrast to the pure polarization case. The injections are conducted for the inclination angle $\iota=\pi/6,\pi/4,\pi/3,\pi/2$. The prior setting for the mixed polarization case is the same as those in the pure polarization case except for the non-GR parameters.

Note that the injected values of the scalar mode amplitude above are too large from the observational point of view when considering modified gravity theories that show the same properties in a strong gravitational field as in a weak gravitational field~\cite{Takeda:2023mhl}.
However, our objective is to examine the effect of inference for the inclination angle on the estimation of the scalar mode amplitude. The qualitative behavior of this effect is independent of the specific value of the scalar mode amplitude. Thus, the following results remain valid even for the smaller amplitude.

\begin{table}[ht]
  \caption{The prior settings. Here, $M_{\odot}$ is the solar mass, and $t_{\mathrm{peak}}=186741861.5$ is the same value as the injected value shown in Table~\ref{tab:injected values for inclination recovery}.}
  \centering
  \begin{tabular}{*{3}{wc{27mm}}}
      \toprule
      \multicolumn{1}{c}{Parameter} & \multicolumn{1}{c}{Prior} & \multicolumn{1}{c}{Range} \\
      \midrule
      $\mathcal{M}~[M_{\odot}]$ & uniform & $[5, 50]$ \\
      $q=m_{2,\mathrm{source}}/m_{1,\mathrm{source}}$ & uniform & $[0.125,1]$ \\
      $\chi_{1,2}$ & delta function & $0$ \\
      $d_{L}~[\mathrm{Mpc}]$ & uniform in the& $[10, 1500]$ \\
      &source frame & \\
      
      $\cos\iota$ & uniform & $[-1,1]$ \\
      $\alpha$ & uniform & $[0,2\pi]$ \\
      $\sin\delta$ & uniform & $[-1,1]$ \\
      $\psi$ & uniform & $[0,\pi]$ \\
      $\phi_{\mathrm{ref}}$ & uniform & $[0, 2\pi]$ \\
      $t_{c}~[\mathrm{s}]$ & uniform & $[t_{\mathrm{peak}}-0.5, t_{\mathrm{peak}}+0.5]$ \\
      $A_{b1}$ & uniform & $[-1, 1]$ \\
      $A_{b2}$ & uniform & $[-1, 1]$ \\
      $\tilde{A}_{b1}$ & uniform & $[-1, 1]$ \\
      $\tilde{A}_{b2}$ & uniform & $[-1, 1]$ \\
      \bottomrule
  \end{tabular}
  \label{tab:priors}
\end{table}
\section{Results}
\label{sec:Results}
\subsection{Inference of inclination angle for pure polarization}
\label{subsec:Inference of inclination angle for pure polarization}
\begin{figure*}[t]
    \begin{tabular}{cc}
      \begin{minipage}[t]{0.48\textwidth}
        \centering
        \includegraphics[keepaspectratio, width=1\linewidth]{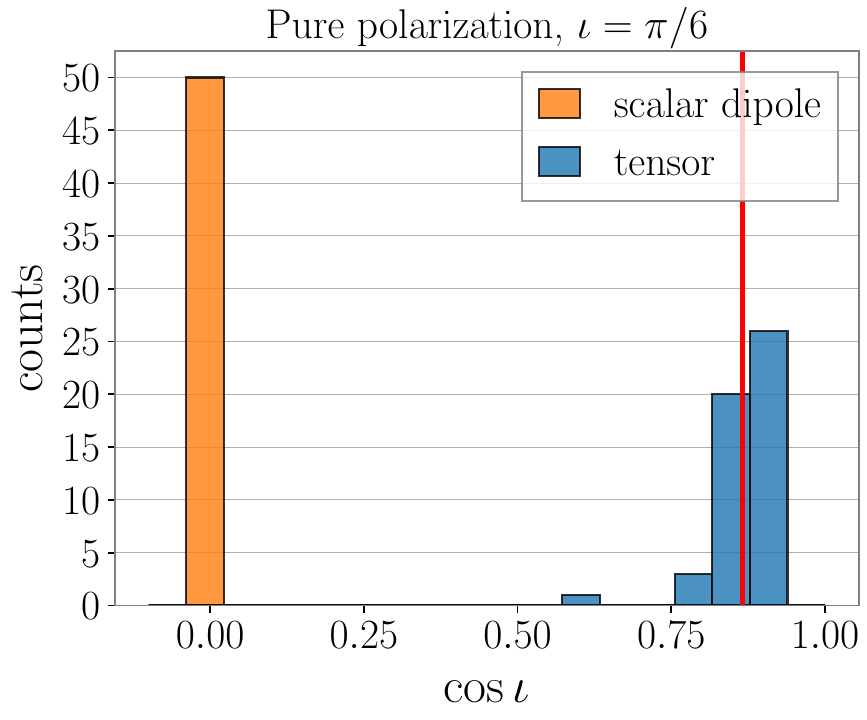}
      \end{minipage} &
      \begin{minipage}[t]{0.48\textwidth}
        \centering
        \includegraphics[keepaspectratio, width=1\linewidth]{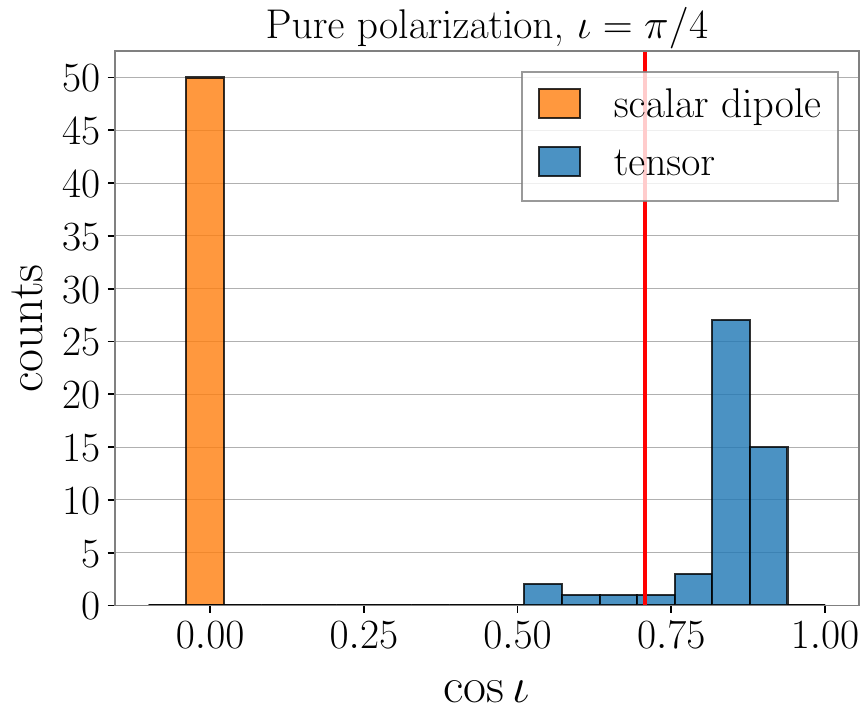}
      \end{minipage} \\
      \begin{minipage}[t]{0.48\textwidth}
        \centering
        \includegraphics[keepaspectratio, width=1\linewidth]{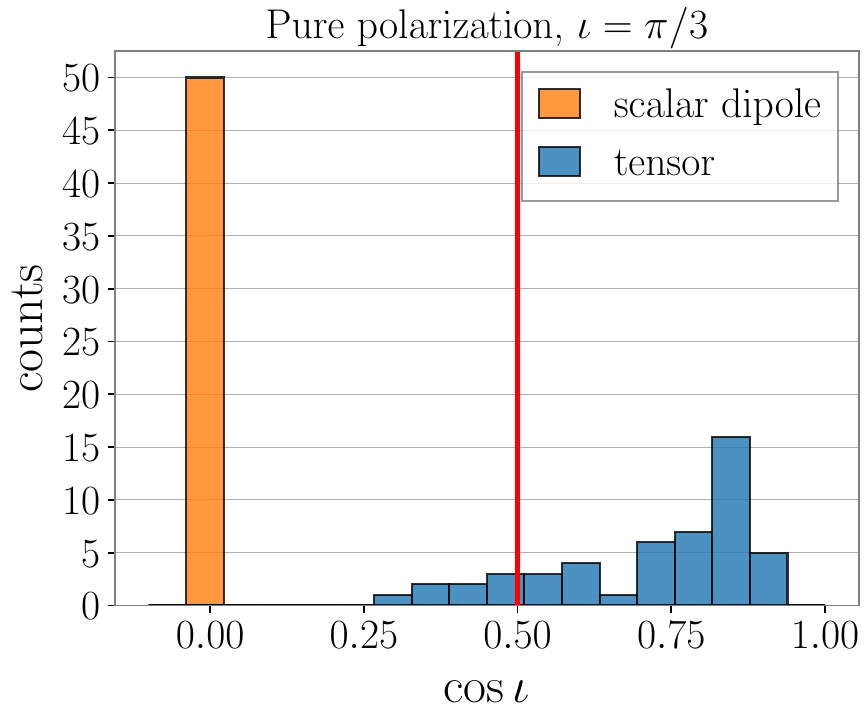}
      \end{minipage} &
      \begin{minipage}[t]{0.48\textwidth}
        \centering
        \includegraphics[keepaspectratio, width=1\linewidth]{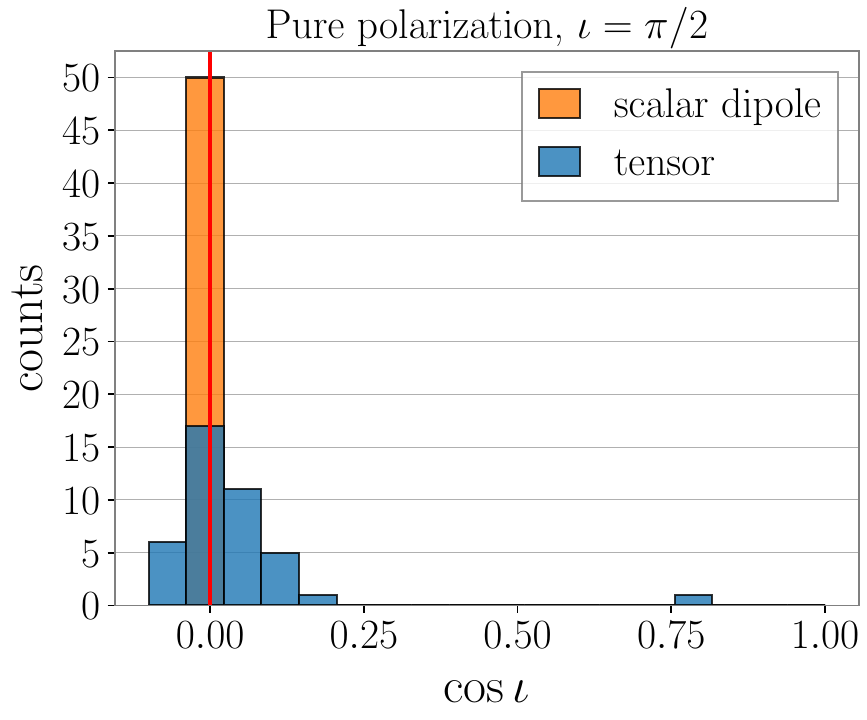}
      \end{minipage} 
    \end{tabular}
    \caption{The histograms of the inclination recovery for the case of pure tensor mode (blue) and pure scalar dipole mode (orange). Each panel shows a different inclination angle, $\pi/6,\pi/4,\pi/3,\pi/2$, from the top left to the bottom right. The red line represents the injection value of the inclination angle.}
    \label{fig:inclination recovery}
\end{figure*}

Figure~\ref{fig:inclination recovery} shows the histograms of the median values of the inclination angle obtained from the posterior probability distribution 
for each injection in the pure polarization case.

In the case of the pure tensor mode, $\cos\iota \sim 1$ is more preferred except when $\iota = \pi/2$. This result can be attributed to two primary factors. First, as mentioned earlier, there is a correlation between the luminosity distance and the inclination angle. Broadly speaking, since the amplitudes of both the plus and cross modes take the same value when $\iota =0$ (or $\pi$), the amplitude of the GR waveform is approximately proportional to $\mathcal{A}=\cos\iota/d_L$, making it challenging to accurately estimate each parameter. Second, we adopt a uniform-in-comoving-volume prior for the luminosity distance, while the sources are assumed to distribute at the same distance. The uniform-in-comoving-volume prior is expressed as $p(d_L|M)\propto d^{2}_{L}$ at low redshift, indicating that larger distances are more likely to be favored. The combination of these two factors influences the estimation of $\cos\iota$, leading to a preference for $\cos\iota\sim1$. This behavior can be understood as follows: the posterior probability of $\cos\iota$ for a given fixed value of $\iota$ is obtained by integrating over the distribution weighted by the prior~\cite{Usman:2018imj},
\begin{align}
    p(\cos\iota) &\propto \cos^3\iota\:.
\end{align}
Therefore, when there is a correlation between the inclination angle and the luminosity distance, the posterior probability of $\cos\iota$ is proportional to $\cos^3\iota$.

However, when $\cos\iota\sim0$, the correlation between the inclination angle and the luminosity distance is 
weak~\cite{Vitale:2014mka, Vitale:2018wlg}. As a result, the inference of the inclination angle (and consequently the luminosity distance) is improved, and the medians are distributed closer to the true value (the bottom right panel in Fig.~\ref{fig:inclination recovery}). Similar results have been reported in other studies~\cite{Usman:2018imj, Chen:2018omi}.

On the other hand, in the case of the pure scalar dipole mode, $\cos\iota\sim0$ is favored.
This is also due to the two factors mentioned earlier. 
The key difference is that the amplitude of the scalar dipole radiation is proportional to $\sin\iota/d_L$. 
From a similar reasoning as in the pure tensor mode case, the posterior probability of $\sin\iota$ is given by
\begin{align}
    p(\sin\iota) \propto \sin^3\iota\:.
\end{align}
Hence, $\iota\sim\pi/2$ is more likely to be estimated.

Furthermore, all results indicate $\cos\iota\sim0$, regardless of the injected inclination angle. 
This occurs because, when the detector network is effectively sensitive to only one polarization, the inclination angle and the luminosity distance are completely degenerated, as the amplitude ratio of each mode cannot be obtained~\cite{Chassande-Mottin:2019nnz}. 
In GR, there are two polarizations, and the number of polarizations to which detectors are sensitive depends on the orientation of the source. However, in the pure scalar case, since there is intrinsically only one polarization, the strong degeneracy remains. Thus, $\cos\iota\sim0$ is strongly favored in all results. 

The results for the pure scalar quadrupole case are not shown here, as they are similar to those for the pure scalar dipole case. 
The amplitude of the scalar quadrupole mode is proportional to $\sin^2\iota/d_L$. 
Consequently, when larger distances are preferred, larger values of $\sin^2\iota$ are also estimated, exhibiting the same qualitative behavior as in the pure scalar dipole case.

The key observation is that in the cases of the pure tensor and the pure scalar modes, opposite values of  $\cos\iota$ are favored: $\cos\iota\sim1$ for the pure tensor mode and $\cos\iota\sim0$ for the pure scalar mode. 
This behavior results from the inclination angle dependence of the amplitude: $\cos\iota/d_L$ and $\sin\iota/d_L$ ($\sin^2\iota/d_L$) for the tensor mode and the scalar dipole (quadrupole) mode, respectively.
In the next section, we will examine how these properties affect the estimation of the scalar mode amplitude in the mixed polarization models.

\subsection{Inference of inclination angle and scalar mode amplitude for mixed polarization}
\label{subsec:Inference of inclination angle and scalar mode amplitude for mixed polarization}
\begin{figure*}[t]
    \begin{tabular}{cc}
      \begin{minipage}[t]{0.48\textwidth}
        \centering
        \includegraphics[keepaspectratio, width=1\linewidth]{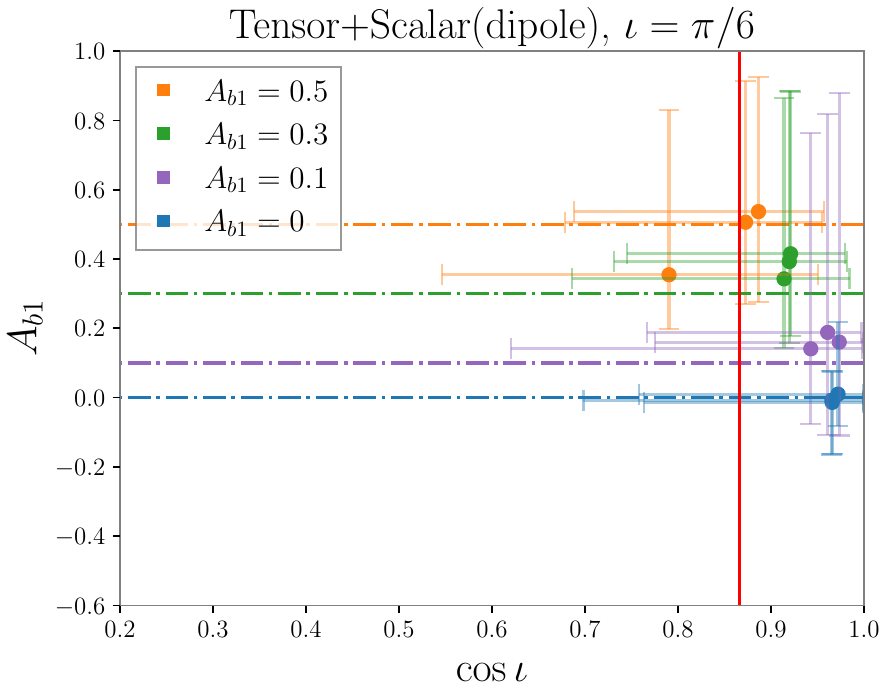}
      \end{minipage} &
      \begin{minipage}[t]{0.48\textwidth}
        \centering
        \includegraphics[keepaspectratio, width=1\linewidth]{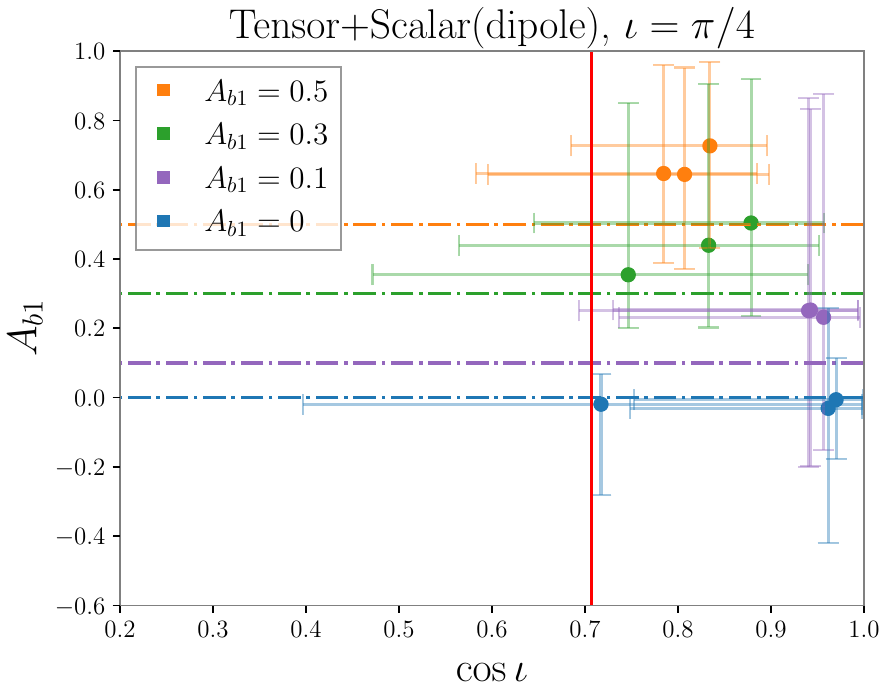}
      \end{minipage} \\
      \begin{minipage}[t]{0.48\textwidth}
        \centering
        \includegraphics[keepaspectratio, width=1\linewidth]{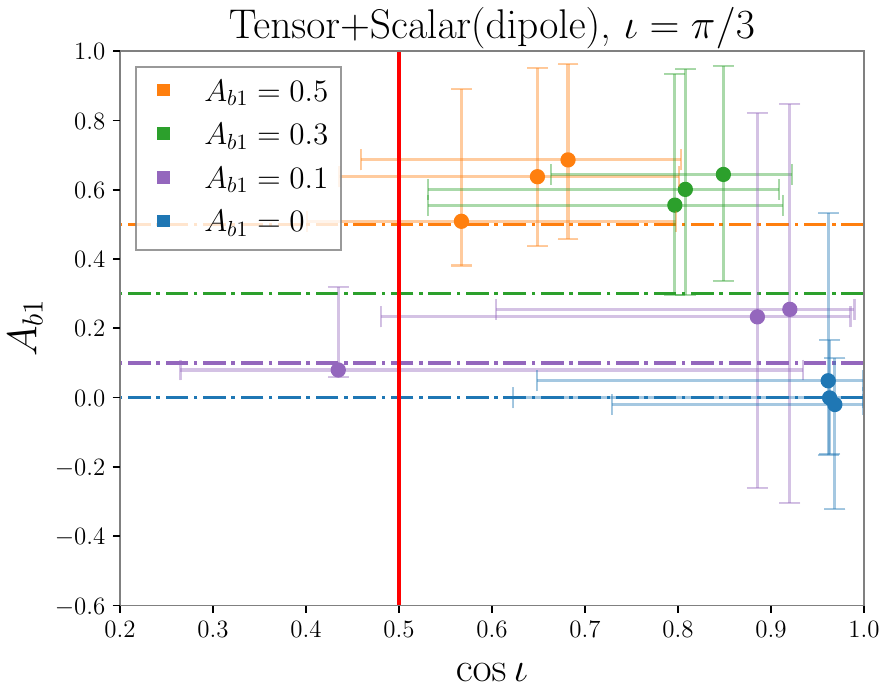}
      \end{minipage} &
      \begin{minipage}[t]{0.48\textwidth}
        \centering
        \includegraphics[keepaspectratio, width=1\linewidth]{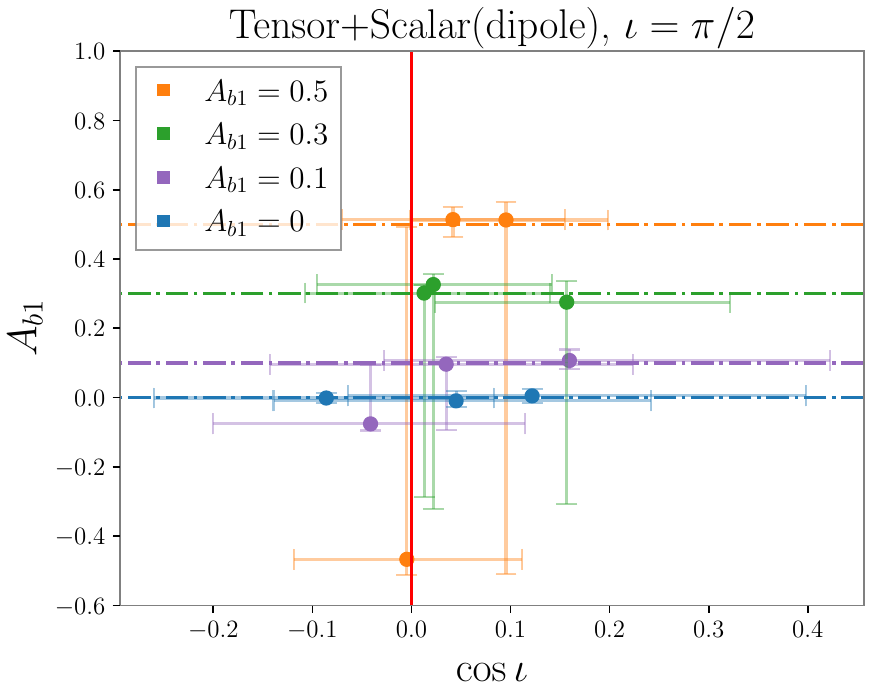}
      \end{minipage} 
    \end{tabular}
    \caption{The medians and the error bars of the inclination angle and the amplitude of the scalar dipole mode. Each panel shows the different injected values of the inclination angle. From top right to bottom left, $\iota=\pi/6,\pi/4,\pi/3,\pi/2$. The colored dots are the medians, and each color corresponds to each injected value of the amplitude of the scalar dipole mode. The vertical red line and horizontal colored line represent the injected value of the inclination angle and the scalar dipole amplitude. The light colored lines indicate the error bars with the $90\%$ credible interval.}
    \label{fig:inclination Ab1 scatter}
\end{figure*}
\begin{figure}[h]
    \centering
    \includegraphics[keepaspectratio, width=1\linewidth]{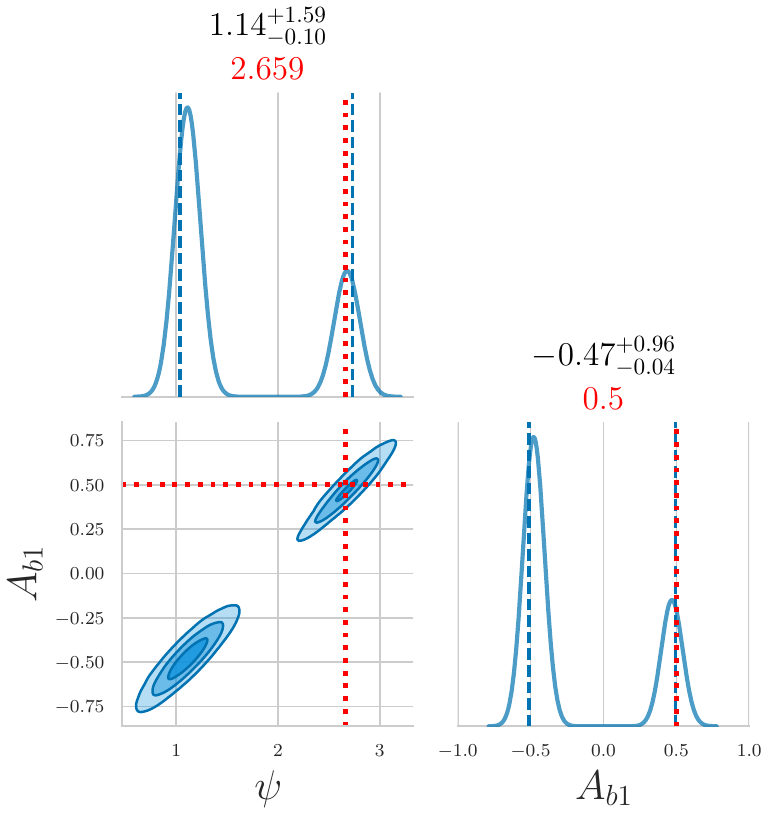}
    \caption{The posterior probability distributions of the polarization angle $\psi$ and the scalar dipole amplitude $A_{b1}$. The red values and lines are the injected values of each parameter. The $1\sigma$, $2\sigma$, and $3\sigma$ contours are shown in the two-dimensional plots. The blue dotted lines represent $90\%$ credible interval in the one-dimensional plots. The values written in black are the medians and errors.}
    \label{fig:Ab1_psi}
\end{figure}
\begin{figure*}[t]
    \begin{tabular}{cc}
      \begin{minipage}[t]{0.48\textwidth}
        \centering
        \includegraphics[keepaspectratio, width=1\linewidth]{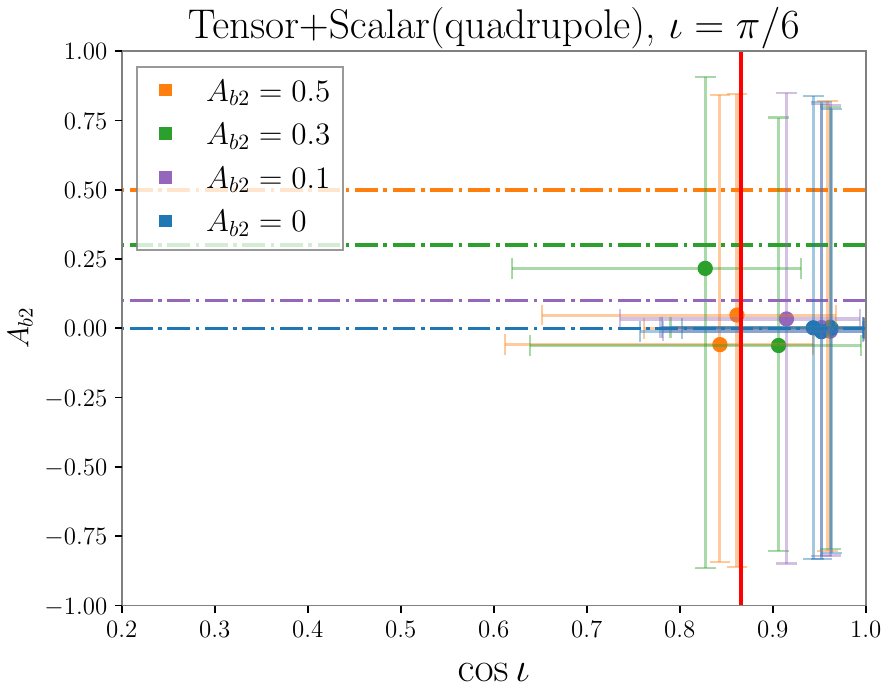}
      \end{minipage} &
      \begin{minipage}[t]{0.48\textwidth}
        \centering
        \includegraphics[keepaspectratio, width=1\linewidth]{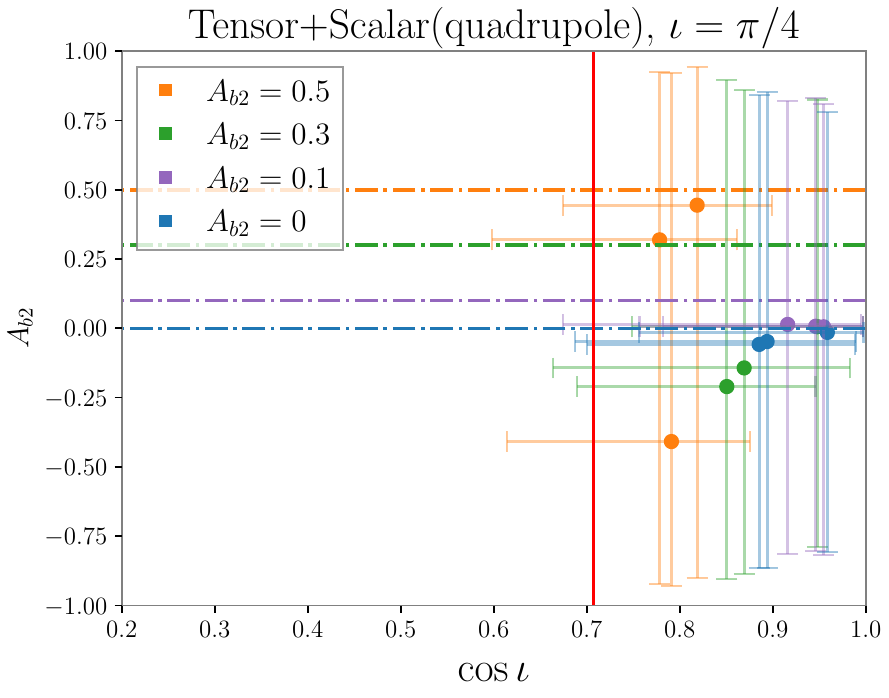}
      \end{minipage} \\
    \end{tabular}
    \centering
    \includegraphics[keepaspectratio, width=0.48\linewidth]{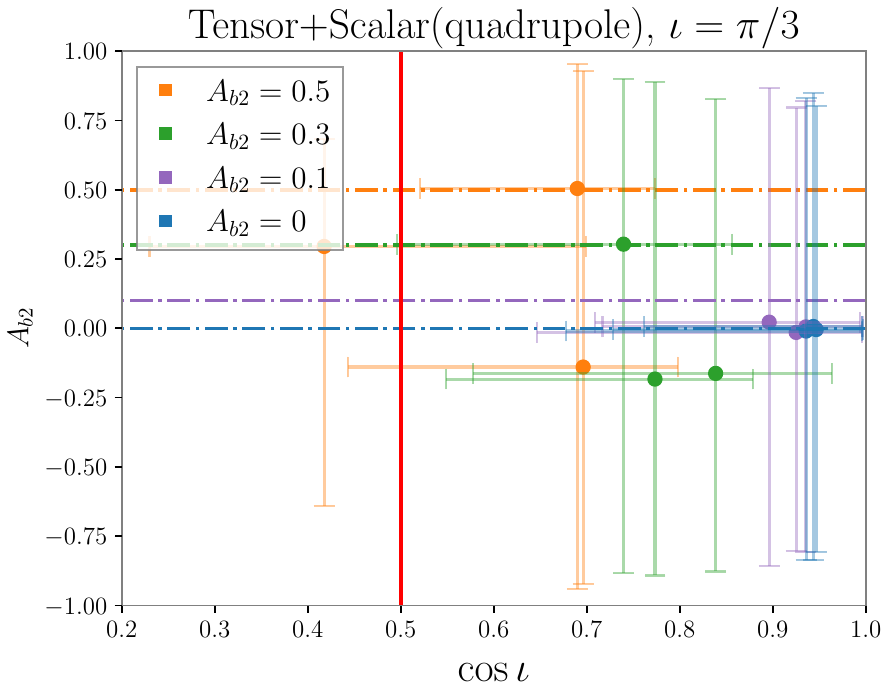}
    \caption{The medians and the error bars of the inclination angle and the amplitude of the scalar quadrupole mode for $\iota=\pi/6,\pi/4,\pi/3$.}
    \label{fig:inclination Ab2 scatter}
\end{figure*}
\begin{figure*}[t]
    \begin{tabular}{cc}
      \begin{minipage}[t]{0.48\textwidth}
        \centering
        \includegraphics[keepaspectratio, width=1\linewidth]{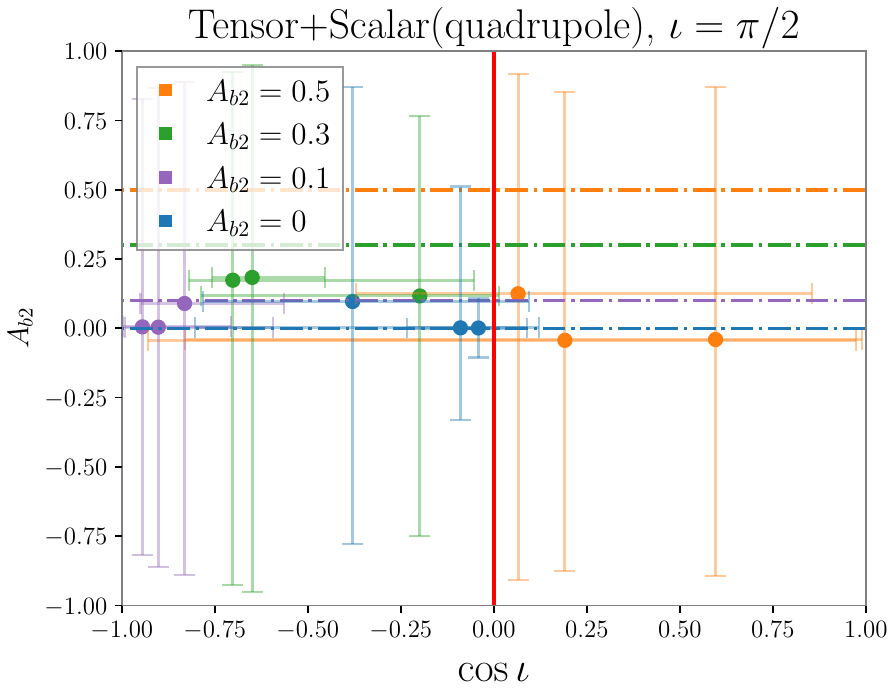}
      \end{minipage} &
      \begin{minipage}[t]{0.48\textwidth}
        \centering
        \includegraphics[keepaspectratio, width=1\linewidth]{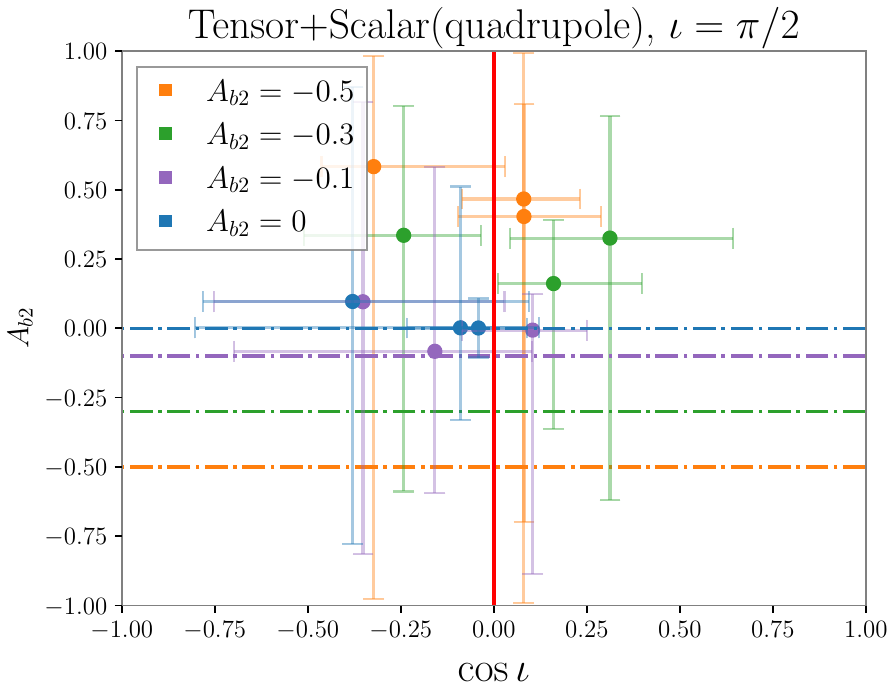}
      \end{minipage} \\
    \end{tabular}
    \caption{The medians and the error bars of the inclination angle and the amplitude of the scalar quadrupole mode for $|A_{b2}|=0,0.1,0.3,0.5$ at $\iota=\pi/2$.}
    \label{fig:inclination Ab2 scatter pi over 2}
\end{figure*}
Figure~\ref{fig:inclination Ab1 scatter} shows the medians and error bars for the inclination angle and the amplitude of the scalar dipole mode. 
Note that for the points with the same color in the same panel, the injection values are identical and the only variation is due to noise realization. 
Focusing first on the measurement of the inclination angle (the horizontal axis), we observe that $\cos\iota$ is estimated more accurately as $A_{b1}$ increases, except when $\iota=\pi/2$.
As discussed in the previous section, the inclination angle preference is opposite for the pure tensor and pure scalar modes.
When $A_{b1}$ is small, the inference is dominated by the tensor mode, resulting in most estimates favoring $\cos\iota\sim1$.
As $A_{b1}$ increases, the scalar dipole mode contributes more significantly to the parameter inference, and pushes the medians of the inclination angle estimated back toward the injected values.
Thus, in the mixed polarization case, the inclination angle estimation reflects a balance between the prior weights on the pure tensor and pure scalar modes.
When $\iota=\pi/2$, the medians of the inclination angle estimates are distributed around the injected value. This is caused by the same reasons as in the pure polarization cases, in which $\iota=\pi/2$ is estimated accurately.

Second, focusing on the estimation of the amplitude of the scalar dipole  mode (the vertical axis), we observe that the amplitude of the scalar dipole emission is overestimated when $A_{b1}$ has a nonzero value, except for $\iota=\pi/2$.
This overestimation arises from the overestimation of $\cos\iota$.
As discussed in the previous paragraph, $\cos\iota\sim1$ is favored when the tensor mode is dominant, leading to an underestimation of $\sin\iota$.
To compensate for this, $A_{b1}$ is overestimated, as the scalar dipole amplitude is expressed as $\tilde{h}_{b1}(f)\propto A_{b1}\sin\iota$.
Notably, this overestimation does not occur when $A_{b1}=0$.
In this case, the injected GW signal contains only the tensor mode, with no scalar dipole radiation.
We can distinguish the presence or absence of the scalar dipole component, preventing false deviations.

At $\iota=\pi/2$, some cases indicate that the median of $A_{b1}$ takes a negative value, or the error bar is large. 
This is attributed to the degeneracy between the polarization angle and the scalar dipole amplitude, which leads to the bimodal posterior distribution for $A_{b1}$, shown in Fig.~\ref{fig:Ab1_psi}. 
When the polarization angle changes as $\psi\rightarrow\psi+\pi/2$, the sign of the tensor mode reverses due to the polarization-angle dependence of the antenna pattern function. 
To maintain the overall signal amplitude, the scalar mode is inferred to have a negative sign.
This phenomenon does not occur when the scalar mode amplitude is relatively small compared to the tensor mode amplitude, as the adjustment of the scalar mode cannot reproduce the true signal. 
$\iota=\pi/2$ makes the tensor mode amplitude smaller and the scalar mode amplitude larger and, therefore, we can see such a property there.

The case of the Tensor+Scalar(quadrupole) model is shown in Figs.~\ref{fig:inclination Ab2 scatter} and \ref{fig:inclination Ab2 scatter pi over 2}.
Most results show that the error bars for the scalar quadrupole amplitude are constrained by the prior range $[-1,1]$.
In this case, since both modes share the same phase evolution, the parameter inference of the scalar mode is more difficult.
Consequently, the results provide little information about $A_{b2}$.

However, regarding the estimation of the inclination angle, we observe similar behavior to that in the Tensor+Scalar(dipole) model: as the scalar mode amplitude increases, the median of the inclination angle approaches the true value in the $\iota\ne\pi/2$ case. This occurs for the same reason as in the Tensor+Scalar (dipole) model. On the other hand, when $\iota=\pi/2$ and the scalar quadrupole amplitude is positive (the left panel of Fig.~\ref{fig:inclination Ab2 scatter pi over 2}), the result is less informative. In this case, the amplitude of the tensor mode becomes small and that of the scalar mode is large due to the dependence on the inclination angle. Furthermore, the sign of the scalar mode amplitude is opposite to that of the tensor mode when $A_{b2}>0$ [see Eqs.~\eqref{FD:full signal}--\eqref{eq:FD quadru}]. These lead to a reduction in the SNRs since the phase evolution of the two modes is the same and they cancel each other. Consequently, the inclination angle is not precisely inferred. To see cases with SNRs comparable to those in other configurations, we inject the negative scalar quadrupole amplitudes of $A_{b2}=0,-0.1,-0.3,$ and $-0.5$ (the right panel of Fig.~\ref{fig:inclination Ab2 scatter pi over 2}). The results show that the medians of the inclination angle are distributed around the true value, as seen in the Tensor+Scalar(dipole) model. This behavior is also due to the tendencies observed in the pure polarization cases.
\section{Discussions and Conclusions}
\label{sec:Discussions and Conclusions}
In this paper, we investigated the effect of the event selection using high-SNR signals on parameter estimation in the scalar-tensor polarization search.
The parametrized scalar-tensor waveform based on Ref.~\cite{Takeda:2023wqn} assumes that the GR waveform is modified due to the changes in the energy loss rate.
This modification is characterized by the non-GR parameters: the coupling parameter $\gamma$, and the amplitude parameters of the scalar dipole mode $A_{b1}$, and the scalar quadrupole mode $A_{b2}$. 
Corrections arising from the scalar mode radiation are incorporated into both the tensor mode and the scalar mode, characterized by the combination of  $\tilde{A}_{b1}=\sqrt{\gamma}A_{b1}$ and $\tilde{A}_{b2}=\sqrt{\gamma}A_{b2}$. 
This waveform model allows for testing GR with a focus on the scalar polarization without assuming any specific theory of modified gravity.

In Sec.~\ref{subsec:Inference of inclination angle for pure polarization}, we analyzed the recovery of the inclination angle for the pure tensor mode and the pure scalar mode when focusing on loud events.
In the pure tensor case, $\cos\iota\sim1$ is favored, whereas $\cos\iota\sim0$ is favored in the pure scalar case.
This behavior arises from two factors: the correlation between the inclination angle and luminosity distance, and the uniform-in-comoving-volume prior on the luminosity distance, which differs from the injected source distribution.
Since the amplitudes of the tensor mode and the scalar mode are proportional to $\cos\iota/d_L$ and $\sin\iota/d_L$, respectively, each mode prefers the opposite side of $\cos\iota$.
The Tensor+Scalar model was investigated in Sec.~\ref{subsec:Inference of inclination angle and scalar mode amplitude for mixed polarization}.
When the amplitude of the scalar dipole mode has a nonzero value, $A_{b1}$ is overestimated due to the underestimation of $\sin\iota$.
In contrast, when $A_{b1}=0$, it is possible to distinguish whether the scalar dipole radiation is present.
As a result, the false deviation does not occur.

In the Tensor+Scalar(quadrupole) model, however, the results for the scalar amplitude measurement were uninformative.
Since both modes share the same phase evolution, there is less sensitivity to $A_{b2}$. 
Regarding inclination angle estimation, we found that $\cos\iota$ is inferred more accurately as the scalar mode amplitude increases in both cases. 
This indicates that the interplay of the characteristics of each pure polarization case determines the accuracy of the inclination angle estimation in the mixed polarization model. 

Our results indicated that performing tests of GR using Bayesian inference with selected GW events can be problematic unless selection effects are properly accounted for in the analysis procedure. 
The mismatch between the source distribution and the prior distribution may lead to biased results.
To address this issue, we propose several approaches to mitigate the selection bias. 
First, conducting statistical searches with multiple GW signals can eliminate the bias introduced by event selection, as the sources will effectively be distributed according to the prior.
Furthermore, this approach is beneficial for improving the parameter estimation sensitivity. 
Given that the scalar mode amplitude in the GW signals is smaller than the tensor mode amplitude, as suggested by the solar system experiments~\cite{Takeda:2023mhl}, statistical methods are particularly crucial to enhance the sensitivity to the scalar polarization modes for the GW signals detected by the second-generation detectors. 

Another approach to mitigate selection bias is to appropriately model the selection process and incorporate it into the prior.
For example, if certain preferences exist for the analyzed GW events, such as those based on the SNR or the detector network, appropriate modeling of the prior would help address the bias. 

On the other hand, when the SNR is sufficiently high, the likelihood dominates over the prior in parameter inference. 
In this case, the analysis results become independent of the prior, allowing for unbiased results even from a single GW signal. 
Notably, the third-generation detectors such as the Einstein Telescope~\cite{Punturo:2010zz} and Cosmic Explorer~\cite{Reitze:2019iox} are expected to detect GW events with SNRs approximately ten times higher than those of the current observed events, enabling more precise and unbiased analyses.
However, since parameter estimation is influenced by various factors other than SNR, such as the source orientation and the number of detectors, it is difficult to determine a definitive SNR threshold for obtaining unbiased results. We leave the quantitative evaluation of the issue for the future work.

In our analysis, only the inspiral phase was considered,  as the constructed scalar-tensor waveform is valid only for this stage. 
The scalar-tensor waveform for the merger and ringdown phases remains poorly understood and challenging to parametrize. 
If it becomes possible to analyze an entire inspiral-merger-ringdown signal within the parametrized framework developed based on the GR inspiral-merger-ringdown waveform~\cite{Watarai:2023yky}, more precise tests of the polarization could be performed, though the precision of the amplitude parameter measurements largely depends on the SNR.
Additionally, we assumed an aligned-spin waveform and neglected  higher-order modes for simplicity. 
Both assumptions influenced the analysis, as higher-order modes and spin misalignment can break the degeneracy between the inclination angle and luminosity distance, e.g.,~\cite{Vitale:2018wlg, London:2017bcn, LIGOScientific:2020stg,Tsutsui:2020bem,Tsutsui:2021izf}, improving the measurement of the scalar mode amplitude. 
Furthermore, our analysis considered a GW detector network consisting of three detectors.
The participation of additional detectors, such as KAGRA~\cite{KAGRA:2020tym, Aso:2013eba, Somiya:2011np} and LIGO India~\cite{Unnikrishnan:2013qwa}, would enable more accurate and precise parameter inference, particularly for the sky localization, the luminosity distance, the inclination angle, and polarizations~\cite{Veitch:2012df, Gaebel:2017zys, Takeda:2018uai}. 
Expanding the detector network would facilitate tests of GR that account for additional GW polarizations, such as the vector-tensor and the scalar-vector-tensor polarizations,
because testing multiple polarizations requires at least the same number of detectors as the number of polarization modes  in principle~\cite{Takeda:2018uai}. 
Therefore, expanding the detector network along with developing statistical methods without selection bias is crucial for searching for additional GW polarizations with high precision and accuracy.

\section*{Acknowledgments}
\label{sec:Acknowledgments}
We would like to thank Hiroyuki Nakano, Kazuya Kobayashi, Kenta Hotokezaka, Masaki Iwaya, Soichiro Morisaki, Stephen Fairhurst, and Tatsuya Narikawa for useful discussions. H.~T. is supported by the Hakubi project at Kyoto University and by Japan Society for the Promotion of Science (JSPS) KAKENHI Grant No.~JP22K14037. A.~N. is supported by JSPS KAKENHI Grants No.~JP23K03408, No.~JP23H00110, and No.~JP23H04893. D.~W. is supported by JSPS KAKENHI Grant No.~23KJ06945.

\appendix
\section{Validity of ra-dec choice for mixed polarization}
\label{sec:Validity of ra-dec choice for mixed polarization}
Here, we assess the validity of the sky location chosen in the mixed polarization case. In the pure polarization cases, GW signals are injected with ra-dec sets sampled from the prior distribution, whereas the sky location with $\alpha=0.833~\mathrm{rad}$ and $\delta=-0.784~\mathrm{rad}$ is used in the mixed polarization case. We confirm that the results are not sensitive to this particular choice.

\begin{figure*}[t]
    \begin{tabular}{cc}
      \begin{minipage}[t]{0.48\textwidth}
        \centering
        \includegraphics[keepaspectratio, width=1\linewidth]{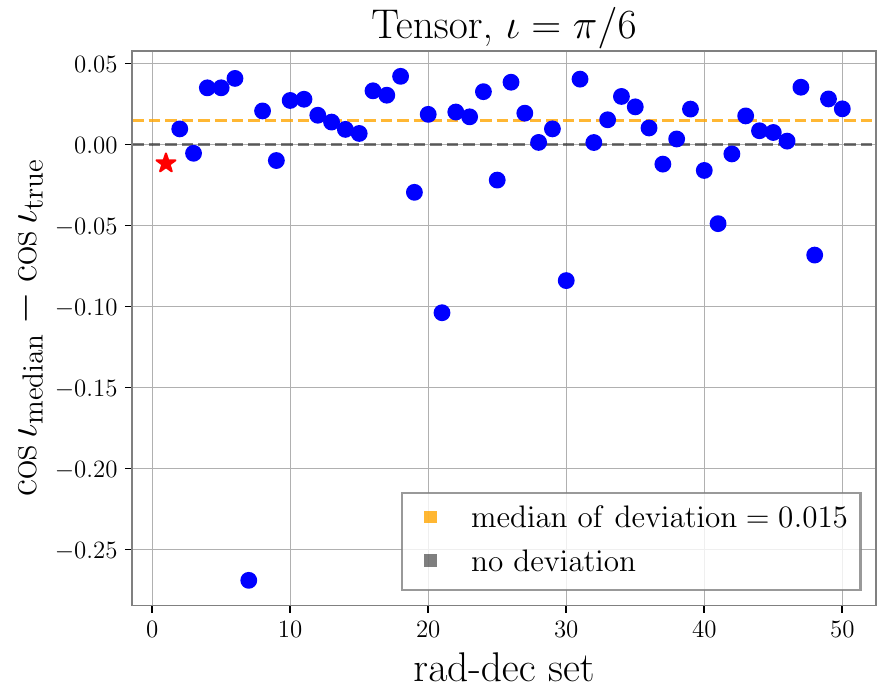}
      \end{minipage} &
      \begin{minipage}[t]{0.48\textwidth}
        \centering
        \includegraphics[keepaspectratio, width=1\linewidth]{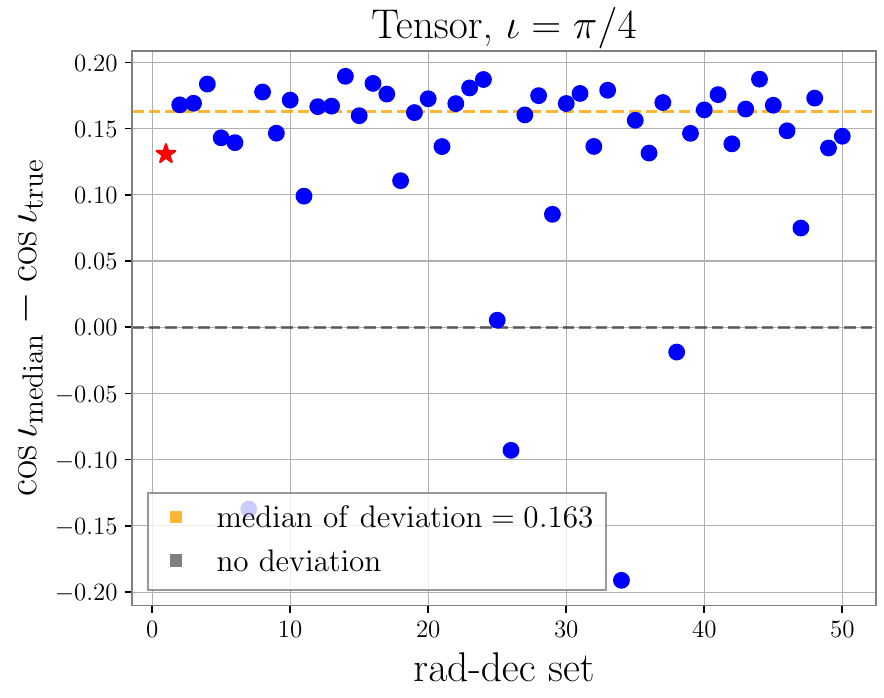}
      \end{minipage} \\
      \begin{minipage}[t]{0.48\textwidth}
        \centering
        \includegraphics[keepaspectratio, width=1\linewidth]{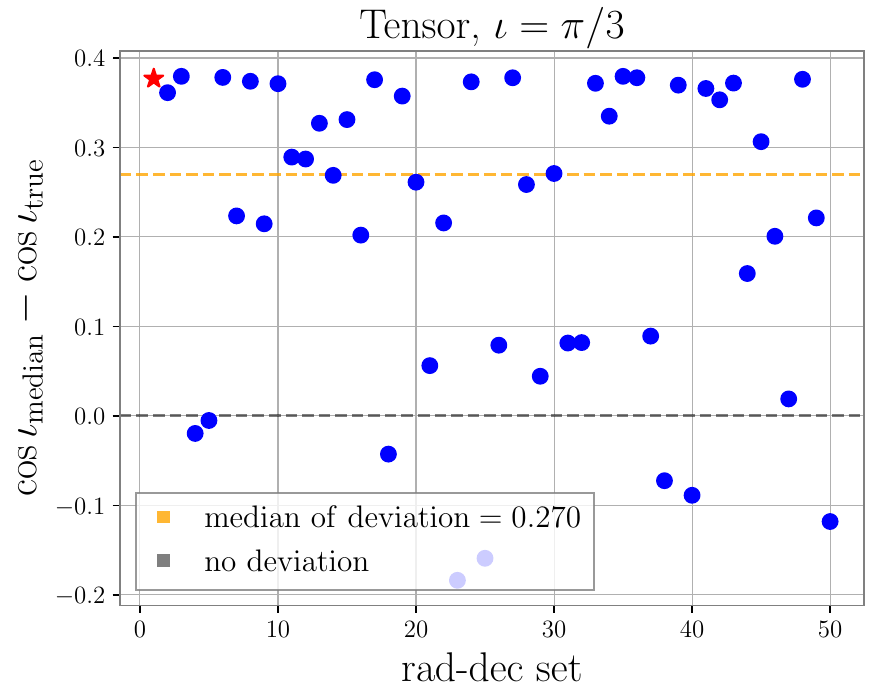}
      \end{minipage} &
      \begin{minipage}[t]{0.48\textwidth}
        \centering
        \includegraphics[keepaspectratio, width=1\linewidth]{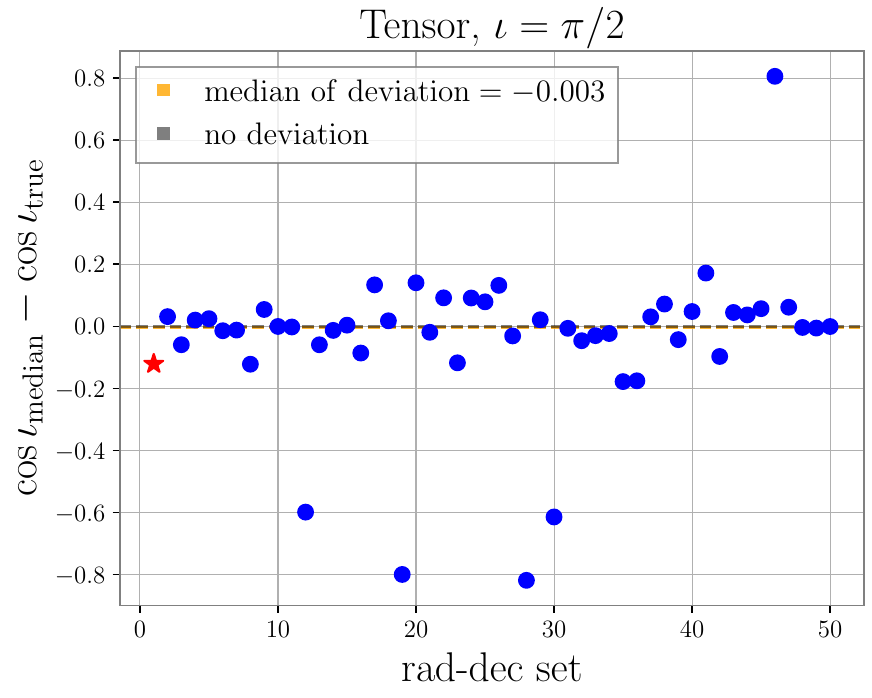}
      \end{minipage} 
    \end{tabular}
    \caption{The dependence of the deviation between the median and the true value of $\cos\iota$, as shown in Fig.~\ref{fig:inclination recovery}, on the sky location in the pure tensor mode case with $\iota = \pi/6, \pi/4, \pi/3, \pi/2$ (from top left to bottom right). The vertical axis shows the difference between the median and the injected value of $\cos\iota$, while the horizontal axis corresponds to the $i$th uniformly random realization of the set of ($\alpha$, $\sin \delta$) used in the mixed polarization case. The orange line represents the median of the deviation across 50 ra-dec sets, and the red star indicates the ra-dec set used in the mixed polarization case.}
    \label{fig:deviation_radec_dependece_tensor}
\end{figure*}

Figure~\ref{fig:deviation_radec_dependece_tensor} represents how the deviation between the median and the true value of $\cos\iota$, as presented in Fig.~\ref{fig:inclination recovery}, depends on the sky location in the pure polarization case. The deviation corresponding to the ra-dec set used in the mixed polarization case, marked with the red star, is not far from the median of the deviation across all ra-dec sets, indicated by the orange line. This shows that the selected ra-dec set does not exhibit any special behavior in terms of the inclination angle estimation.
Note that we only present the pure tensor case here, as the deviations in the pure scalar case are nearly uniform across all sky locations.

\bibliographystyle{apsrev4-2}
\bibliography{main}

\end{document}